\documentclass[11pt,a4paper]{article}

\usepackage{lineno,hyperref}
\usepackage{graphics,color,cite}
\usepackage{amsmath}
\usepackage{latexsym,amsmath,amssymb,amsfonts,amsthm,amsbsy,enumerate}
\usepackage{amsmath,amssymb}
\usepackage[affil-it]{authblk}

\begin{document}

\title{Coherent-state path integrals in the continuum:\\ The $SU(2)$ case}

\author[1]{G. Kordas \thanks{gekordas@phys.uoa.gr}}

\author[1]{D. Kalantzis}

\author[1]{A.~I. Karanikas}

\affil[1]{University of Athens, Physics Department,Panepistimiopolis, Ilissia 15771 Athens, Greece}

\maketitle

\begin{abstract}
We define the time-continuous spin coherent-state path integral in a way that is free from inconsistencies. The proposed definition is used to reproduce known exact results. Such a formalism opens new possibilities for applying approximations with improved accuracy and can be proven useful in a great variety of problems where spin Hamiltonians are used.
\end{abstract}

\section{\label{sec:I}Introduction}

The Feynman path integral formalism~\cite{Feynman48,Feynman51} provides the most powerful tool for taking into account quantum behavior via classical computations. Ideally suited for semi-classical calculations the path integral machinery helps handling and understanding quantum mechanics, quantum field theories or statistical physics~\cite{Kleinert}.

The extension of path integration into the ordinary complex plane $C^1$, through the Glauber coherent-states~\cite{Glauber63}, and in the complex compact non-flat manifold $\bar{C}^1$, through the $SU(2)$ spin coherent-states~\cite{Klauder77,Berezin80}, has expanded its range of applications in many areas of physics and chemistry~\cite{Kleinert,Klauder85}. The details of these extensions and their utility for semi-classical approximations have been discussed in a lot of excellent papers~\cite{Klauder79,Kurchan79,Solari87,Kohetov95,Kohetov98,Baranger01,Garg03}. In almost all of them the authors point out the fact that path integrals in the continuum are formal limiting expressions of an underlying discrete definition meaning that, in case of discrepancies, one must refer to the discrete version of the integrals. For example, the spin-coherent state path integrals were thought to be unreliable in their continuous form and trustable only in their discretized form~\cite{Karatsuji81,Carg92,Enz86,Enz86_2,Belinicher97,Shibata99}. It was only after the emergence in the continuum~\cite{Kohetov98,Garg03} of the Solari's~\cite{Solari87} ``extra phase'' that the try for a trustable continuous formulation of the spin-coherent state path integration has been renewed.

Nevertheless however, the problems still persist as inconsistencies and wrong results have been reported recently~\cite{Wilson11} even for simple and exactly solvable systems when examined via coherent-state path integrals in the continuum. Interesting enough, even with the inclusion of the ``extra phase'' contribution, a system described by a Hamiltonian of the form $\hat{H}\sim \hat{S}_z^2$ cannot be described correctly in terms of path integration in the continuum.

However, after the extensive use of the continuous formulation in almost all the fields of quantum theory~\cite{Kleinert} or after the Berezin's quantization scheme for non-flat manifolds like $\bar{C}^1$~\cite{Berezin80}, it would be at least awkward if it was to be concluded that, in the framework of the coherent state path integrals, it is impossible to define a classical continuous action in correspondence with a quantum system.

In a recent paper~\cite{Kordas14} we examined the case of the continuous formulation of path integrals in terms of the Glauber coherent states. We found that the inconsistencies disappear if one follows a certain recipe to define the classical Hamiltonian entering in the continuous action that weighs the paths in the complex plane. In the present work we extend our work undertaking the task of establishing a connection between
the quantum Hamiltonian and the continuous action appropriate for path
integration in the spin coherent-states basis and we discuss some aspects of the corresponding time sliced definition. In the context of the proposed formulation the path integration can be performed directly in the continuum without facing inconsistencies and reproduces the exact results.

The paper is organized as follows. In Sec.~\ref{sec:II} we present our proposal in the context of the simplest possible system $\hat{H}_1=\omega\hat{S}_z$ which can be exactly (and correctly) handled by a lot of means. We examine this system in the framework of the ``standard'' spin coherent state path integration and we compare the result with the analysis based on our proposal. In Sec.~\ref{sec:III} we consider the case of a system described by a Hamiltonian of the form $\hat{H}=\omega\hat{S}_z^2$ for the description of which the standard approach breaks down. We prove that in the framework of our proposal the spin coherent-state path integration yields the correct answer. In the last section we summarize our findings and we comment on possible applications.

\section{\label{sec:II}A simple example}

To present our arguments we begin by considering the simplest possible case: a single particle with spin $s$ in an external constant magnetic field. The Hamiltonian of such a system is $\hat{H}_1=\omega\hat{S}_z$ and the representation of the time evolution operator in the spin coherent state basis reads:
\begin{equation}\label{eq:01}
G_1(\zeta_b^*,\zeta_a) = \langle\zeta_b|e^{-iT\hat{H}_1}|\zeta_a\rangle.
\end{equation}
The states of this basis can be defined~\cite{Radcliff71} through the relation
\begin{eqnarray}\nonumber
|\zeta\rangle &=& \frac{1}{\left(1+|\zeta|^2\right)^s} e^{\zeta \hat{S}_-}|s,s\rangle\\
 &=& \frac{1}{\left(1+|\zeta|^2\right)^s} \sum_{j=-s}^s \left[ \frac{(2s)!}{(s-j)!(s+j)!}\right]^{1/2} \zeta^{s-j} |s,j\rangle,\label{eq:02}
\end{eqnarray}
where $|s,s\rangle$ is the eigenstate of $\hat{S}_z$ with the largest eigenvalue. The states (\ref{eq:02}) form an overcomplete basis on the compact non-flat manifold $\bar{C}^1$, the one-point compactified complex plane, that is identified with the $SU(2)$ homogeneous space, $SU(2)/U(1)$~\cite{Kohetov95,Kohetov98}:
\begin{eqnarray}
\hat{I} &=& \frac{2s+1}{\pi}\int\frac{d^2\zeta}{\left(1+|\zeta|^2\right)^2},\label{eq:3a}\\
\int d^2\zeta &\equiv& \int d\Re \zeta \int d\Im \zeta.\label{eq:3b}
\end{eqnarray}
This resolution of the identity can be used to present matrix elements like the amplitude (\ref{eq:01}) as integrals over paths in the complex space $\bar{C}^1$~\cite{Kohetov95,Kohetov98}:
\begin{equation}\label{eq:04}
G(\zeta_b^*,\zeta_a) = \mathop{\int \mathcal{D}\mu(\zeta)}\limits_{\substack{\zeta^*(T)=\zeta_a^* \\ \zeta(0)=\zeta_a}} e^{\gamma_{ba}(\zeta^*,\zeta) + iS[\zeta^*,\zeta]}.
\end{equation}
In this expression
\begin{eqnarray}\label{eq:05}
\int \mathcal{D} \mu(\zeta)~(\cdots) &\equiv& \lim_{N\rightarrow\infty} \left(\prod_{n=1}^{N-1} \frac{2s+1}{\pi}\int \frac{d^2\zeta_n}{\left( 1+ |\zeta|^2\right)^2} \right)~(\cdots),\\
\gamma_{ba} &=& s \ln\frac{(1+\zeta_b^* \zeta(T))(1+\zeta_a\zeta^*(0))}{\left( 1+ |\zeta_b|^2\right)\left( 1+ |\zeta_a|^2\right)}
\end{eqnarray}
and
\begin{equation}\label{eq:07}
S = \int_0^T d\tau \left( i s \frac{\dot{\zeta}\zeta^* - \dot{\zeta}^*\zeta}{1+|\zeta|^2} - H(\zeta^*,\zeta)\right).
\end{equation}
The classical Hamiltonian in the action (\ref{eq:07}) is supposed to have the following ``standard'' form
\begin{equation}\label{eq:08}
H(\zeta^*,\zeta) = \langle \zeta | \hat{H}|\zeta\rangle.
\end{equation}
For the case in hand one easily finds that
\begin{equation}\label{eq:09}
H_1 = \omega s\left( 1 - 2\frac{|\zeta|^2}{1+|\zeta|^2} \right).
\end{equation}
Thus
\begin{equation}\label{eq:10}
G(\zeta_b^*,\zeta_a) = \mathop{\int \mathcal{D}\mu(\zeta)}\limits_{\substack{\zeta^*(T)=\zeta_a^* \\ \zeta(0)=\zeta_a}} e^{\gamma_{ba}(\zeta^*,\zeta) + i\int_0^T d\tau \left( i s \frac{\dot{\zeta}\zeta^* - \dot{\zeta}^*\zeta}{1+|\zeta|^2} + 2\omega s \frac{|\zeta|^2}{1+|\zeta|^2}\right)}.
\end{equation}
The functional integral appearing in the last equation can be exactly evaluated directly in the continuum~\cite{Kohetov95,Kohetov98,Garg03}.
The procedure goes as follows. Firstly, one finds the ``classical'' solutions pertaining to the action (\ref{eq:07}): $\zeta_c(\tau) = \zeta_a e^{i\omega\tau},~\zeta_c^*(\tau) = \zeta_b^* e^{i\omega(T-\tau)}.$ Then, the change of variables $\zeta = \zeta_c + \delta\zeta,~\zeta^* = \zeta^*_c + \delta\zeta^*$ leads to a prefactor together with a determinant that encapsulates quantum fluctuations and which can be written in terms of a functional integral over complex variables $\eta$:
\begin{eqnarray}
\nonumber G_1(\zeta_b^*,\zeta_a) &=& e^{-i\omega T s} \frac{(1 + \zeta_b^* \zeta_a e^{i\omega T})^{2s}}{(1+|\zeta_b|^2)^s(1+|\zeta_a|^2)^s}\times \\
\label{eq:11} && \times \mathop{\int \mathcal{D}^2\eta} \limits_{\substack{\eta^*(T)=0 \\ \eta(0)=0}} e^{i\int_0^T d\tau \left( \frac{i}{2}(\dot{\eta}\eta^* - \dot{\eta}^*\eta)+\omega|\eta|^2\right)}.
\end{eqnarray}
The fluctuating integral must be carefully evaluated because the result strongly depends~\cite{Kleinert} on the underlying discrete prescription that defines the continuum version appearing in Eq. (\ref{eq:11}). To be concrete, let' s consider the following discrete version of this integral:
\begin{eqnarray}\label{eq:12}
 I &=& \lim_{N\rightarrow\infty}\left(\prod_{n=1}^{N-1}\int\frac{d^2\eta_n}{\pi} \right)\times \\
\nonumber &&\times \exp\Bigg\{\sum_{n=0}^{N-1} \Bigg[\frac{1}{2}[(\eta_{n+1}^* - \eta_n^*)\eta_n - (\eta_{n+1}-\eta_n)\eta_{n+1}^*]
 +i\varepsilon\omega\eta_{n+1}^*\eta_n\Bigg]\Bigg\},
\end{eqnarray}
with $\varepsilon = T/N$. This integral can be straightforwardly evaluated~\cite{Kleinert} and the result is $I=1$.

Roughly speaking , this result is connected with the possibility to make a change of variables $\eta\rightarrow\eta e^{i\omega\tau},~\eta^*\rightarrow\eta^* e^{-i\omega\tau}$ in the continuum level or $\eta_n \rightarrow \eta_n(1+i\omega\varepsilon n),~\eta_n^* \rightarrow \eta_n^*(1-i\omega\varepsilon n)$ in the discrete level. This change leaves intact the measure of the integration but not the action (in this sense it constitutes an ``anomaly'')~\cite{Garg03} as it cancels the term $\omega |\eta|^2$ which has been taken to be the continuum limit of the term $\varepsilon\omega\eta_{n+1}^*\eta_n$ in the discretized version of the action.

Consider now the following ``symmetrized'' definition of the fluctuating path integral:
\begin{eqnarray}\label{eq:13}
 I_S &=& \lim_{N\rightarrow\infty}\left(\prod_{n=1}^{N-1}\int\frac{d^2\eta_n}{\pi} \right)\times \\
\nonumber &&\times \exp\Bigg\{\sum_{n=0}^{N-1} \Bigg[\frac{1}{2}[(\eta_{n+1}^* - \eta_n^*)\eta_n - (\eta_{n+1}-\eta_n)\eta_{n+1}^*] 
 + i\varepsilon\omega\eta_{n}^*\eta_n\Bigg]\Bigg\},
\end{eqnarray}
where $\varepsilon = T/N$.Despite the fact that the continuum limit of the two expressions is the same the result of the calculation is different~\cite{Kleinert}: $I_S=e^{i\omega T/2}$. In the formal level, one can check that the previously discussed change of variables does not cancel the quadratic term in the symmetrized action. Leaving for later the discussion about which of the mathematically possible approaches is the physically correct one, we shall adopt here the version (\ref{eq:12}). In this case we find:
\begin{equation}\label{eq:14}
G_1(\zeta_b^*,\zeta_a) = e^{-i\omega T s} \frac{(1 + \zeta_b^* \zeta_a e^{i\omega T})^{2s}}{(1+|\zeta_b|^2)^s(1+|\zeta_a|^2)^s}.
\end{equation}
By taking the trace of the amplitude (\ref{eq:14}) we can immediately confirm that the correct result is produced:
\begin{eqnarray}\label{eq:15}
\nonumber \frac{2s+1}{\pi} \int \frac{d^2\zeta}{(1+|\zeta|^2)^2)} G_1(\zeta^*,\zeta) &=& e^{-i\omega T s}\sum_{p=0}^{2s} e^{i\omega T p} \\
&=& \sum_{j=-s}^s e^{-i\omega T j} = {\rm Tr}\{ e^{-iT\hat{H}_1}\}.
\end{eqnarray}
A warning for a possible pitfall in the above described procedure, and especially in the form of the classical Hamiltonian, comes up~\cite{Wilson11} when one starts to consider more complicated systems. As we shall discuss in the next section, the case of the simple but less trivial Hamiltonian $\hat{H}_2 = \omega\hat{S}_z^2$ cannot be correctly analyzed through this formulation.

Our next step is, now, to consider the same system at the limit $s\rightarrow\infty$. At this limit the spin coherent states (\ref{eq:02}) reduce~\cite{Kohetov98,Radcliff71} to the harmonic oscillator coherent states:
\begin{equation}\label{eq:16}
|\zeta\rangle \xrightarrow[s\rightarrow\infty]{} e^{-\frac{1}{2}|z|^2}\sum_{n=0}^\infty \frac{z^n}{\sqrt{n!}}|0\rangle,~(\zeta\rightarrow z/\sqrt{2s}).
\end{equation}
At the same limit the functional integral reduces to a functional integral over the states (\ref{eq:16}):
\begin{equation}\label{eq:17}
G(\zeta_b^*,\zeta_a) \rightarrow G(z_b^*,z_a) = \mathop{\int \mathcal{D}^2 z} \limits_{\substack{z^*(T)=z_b^* \\ z(0)=z_a}} e^{\Gamma_{ba} + iS[z^*,z]}.
\end{equation}
Here
\begin{eqnarray}
\label{eq:18a}\int\mathcal{D}^2 z (\cdots) &\equiv& \lim_{N\rightarrow\infty} \left( \prod_{n=1}^{N-1} \int \frac{d^2 z_n}{\pi}\right) (\cdots),\\
\label{eq:18b} \Gamma_{ba} &=& -\frac{1}{2} \left(|z_b|^2 + |z_a|^2\right)
 + \frac{1}{2} (z_b^* z(T) + z_a z^*(0))
\end{eqnarray}
and
\begin{equation}\label{eq:19}
S=\int_0^T d\tau \left(\frac{i}{2}(\dot{z}z^* -\dot{z}^* z) - H(z^*,z)\right).
\end{equation}
As we have discussed in~\cite{Kordas14}, the identification of the classical Hamiltonian in this continuous expression is not a trivial task: In order to correctly perform calculations in the continuum, one must follow a certain route to define the classical action that enters in the path integral. To describe how this can be realized in the present case, we express the quantum Hamiltonian, $\hat{H}_1$, of our system in terms of the harmonic oscillator creation and annihilation operators by making use of the Holstein-Primakoff transformation~\cite{Holstein40}:
\begin{equation}\label{eq:20}
\hat{S}_z = s - \hat{\alpha}^\dag \hat{\alpha},~\hat{S}_+ = \sqrt{2s - \hat{\alpha}^\dag \hat{\alpha}}\hat{\alpha},~\hat{S}_- = \hat{\alpha}^\dag \sqrt{2s - \hat{\alpha}^\dag \hat{\alpha}}.
\end{equation}
Using as starting point the quantum Hamiltonian, $\hat{H} = H(\hat{\alpha}^\dag,\hat{\alpha})$ our proposal proceeds with the quadratures $\hat{q} = (\hat{\alpha}^\dag + \hat{\alpha})/\sqrt{2},~\hat{q} = i(\hat{\alpha}^\dag - \hat{\alpha})/\sqrt{2}$ to produce the ``Feynman'' Hamiltonian $\hat{H}_F = H_F(\hat{p},\hat{q})$. Then, the classical Hamiltonian that must weigh the paths in the complex space spanned by the coherent states $\{|z\rangle\}$, is obtained by taking the classical counterpart of $\hat{H}_F$:
\begin{equation}\label{eq:21}
\hat{H}_F \rightarrow H_F(p,q).
\end{equation}
In this expression the classical ``momentum'' and ``position'' variables are defined through the following canonical transformation:
\begin{eqnarray}
\label{eq:22a} p &\equiv& \langle z|\hat{p}|z\rangle =\frac{i}{\sqrt{2}}(z^* - z),\\
\label{eq:22b} q &\equiv& \langle z|\hat{q}|z\rangle =\frac{1}{\sqrt{2}}(z^* + z).
\end{eqnarray}
For the case in hand we have
\begin{eqnarray}
\nonumber\hat{H}_1 = \omega(s - \hat{\alpha}^\dag \hat{\alpha}) &\rightarrow& \hat{H}_{1F} = \omega \left[ s-\frac{1}{2}(\hat{p}^2+\hat{q}^2)+\frac{1}{2}\right] \rightarrow \\
\nonumber&\rightarrow& H_{1F} = \omega \left[ s-\frac{1}{2}(p^2+q^2)+\frac{1}{2}\right]\\
\label{eq:23} && \phantom{H_{1F}} = \omega\left( s - |z|^2 + \frac{1}{2}\right).
\end{eqnarray}
In this way the correlator (\ref{eq:17}) assumes the form:
\begin{eqnarray}\label{eq:24}
 G_{1F}(z_b^*,z_a) &=& \mathop{\int \mathcal{D}^2 z} \limits_{\substack{z^*(T)=z_b^* \\ z(0)=z_a}} e^{\Gamma_{ba} + i\int_0^T d\tau \left(\frac{i}{2}(\dot{z}z^* - \dot{z}^*z) - H_{1F}\right)}\\
\nonumber &=& e^{-i\omega T s - \frac{i}{2}\omega T} \mathop{\int \mathcal{D}^2 z} \limits_{\substack{z^*(T)=z_b^* \\ z(0)=z_a}} e^{\Gamma_{ba} + i\int_0^T d\tau \left(\frac{i}{2}(\dot{z}z^* - \dot{z}^*z) + \omega|z|^2\right)}.
\end{eqnarray}
The change of variables
\begin{eqnarray}
\label{eq:25a} z(\tau) &=& z_a e^{i\omega\tau} + \eta(\tau),\\
\label{eq:25b} z^*(\tau) &=& z_b^* e^{i\omega(T-\tau)} + \eta^*(\tau)
\end{eqnarray}
reduces the amplitude (\ref{eq:24}) into the form:
\begin{eqnarray}
\nonumber G_{1F}(z_b^*,za) &=& e^{-i\omega T s - \frac{i}{2}\omega T} e^{-\frac{1}{2}(|z_b|^2 + |z_a|^2) + z_b^* z_a e^{i\omega T}}\times\\
\label{eq:26}&& \times \mathop{\int \mathcal{D}^2 \eta} \limits_{\substack{\eta^*(T)=0 \\ \eta(0)=0}} e^{i\int_0^T d\tau \left(\frac{i}{2}(\dot{\eta}\eta^* - \dot{\eta}^*\eta) + \omega|\eta|^2\right)}.
\end{eqnarray}
The functional integral in the last equation is the same as the one appearing in Eq.(\ref{eq:11}) but now we have the necessary information to decide which one of the discretized definitions (\ref{eq:12}) or (\ref{eq:13}) is physically correct~\cite{Kleinert}. Since the classical action in the functional integrals (\ref{eq:24}) or (\ref{eq:26}) arose from the oscillator action by means of the canonical transformations (\ref{eq:22a}), (\ref{eq:22b}) and since the classical dynamics are invariant under canonical transformations, the associated amplitudes must be the same:
\begin{eqnarray}
\nonumber\mathop{\int \mathcal{D}^2 \eta} \limits_{\substack{\eta^*(T)=0 \\ \eta(0)=0}} e^{i\int_0^T d\tau \left(\frac{i}{2}(\dot{\eta}\eta^* - \dot{\eta}^*\eta) + \omega|\eta|^2\right)} &=& \langle 0| e^{\frac{i}{2}T\omega(\hat{p}^2 + \hat{q}^2)} |0\rangle \\
\nonumber &=& \langle 0| e^{iT\omega \hat{\alpha}^\dag \hat{\alpha}}|0\rangle e^{\frac{i}{2}T\omega}\\
\label{eq:27} &=& e^{\frac{i}{2}T\omega}.
\end{eqnarray}
This result fixes the physically correct time-slicing to the symmetric one (see Eq.(\ref{eq:13})). Consequently
\begin{equation}\label{eq:28}
G_{1F}(z_b^*,z_a) = e^{i\omega T s} e^{-\frac{1}{2}(|z_b|^2 + |z_a|^2) + z_b^* z_a e^{i\omega T}}.
\end{equation}
By taking the trace of this expression we arrive at the correct result:
\begin{equation}\label{eq:29}
{\rm Tr} (e^{-iT\hat{H}_1}) \stackrel{s\rightarrow\infty}{=} {\rm Tr}~G_{1F} = e^{-i\omega T s} \left( \sum_{p=0}^{2s} e^{i\omega T p}\right) = \sum_{j=-\infty}^\infty e^{-i\omega T j}.
\end{equation}
It is obvious now that this analysis raises a question about the validity of the result (\ref{eq:14}): Suppose that we take the limit $s\rightarrow\infty$ of the integral appearing in Eq. (\ref{eq:10})
\begin{eqnarray}
\nonumber && \mathop{\int \mathcal{D} \mu(\zeta)} \limits_{\substack{\zeta^*(T)=\zeta_b^* \\ \zeta(0)=\zeta_a}} e^{\gamma_{ba}(\zeta^*,\zeta) + i\int_0^T d\tau \left( is\frac{\dot{\zeta}\zeta^* - \dot{\zeta}^* \zeta}{1+|\zeta|^2} + 2s\omega \frac{|\zeta|^2}{1+|\zeta|^2}\right)} \xrightarrow[s\rightarrow\infty]{}\\
 && \mathop{\int \mathcal{D}^2 z} \limits_{\substack{z^*(T)=z_b^* \\ z(0)=z_a}} e^{\Gamma_{ba} + i\int_0^T d\tau \left(\frac{i}{2}(\dot{z}z^* - \dot{z}^* z) + \omega|z|^2 \right)}
 = \langle z_b| e^{\frac{i}{2}\omega T(\hat{p}^2 + \hat{q}^2)}|z_a\rangle.\label{eq:30}
\end{eqnarray}
The last amplitude can be easily calculated in configuration space:
\begin{eqnarray}
\nonumber \langle z_b| e^{\frac{i}{2}\omega T(\hat{p}^2 + \hat{q}^2)}|z_a\rangle &=& \int dx \int dx' \langle z_b|x\rangle \langle x| e^{\frac{i}{2}\omega T(\hat{p}^2 + \hat{q}^2)}|x'\rangle\langle x'|z_a\rangle \\
&=& e^{-\frac{1}{2}(|z_b|^2 + |z_a|^2) + z_b^* z_a e^{i\omega T} + \frac{i}{2}\omega T}.\label{eq:31}
\end{eqnarray}
Obviously the limit $s\rightarrow\infty$ of the result (\ref{eq:11}) does not coincide, as it should, with the last conclusion except if the symmetrized version is adopted. On the other hand , if we adopt this version the calculation of the correlator (\ref{eq:04}) yields the result
\begin{equation}\label{eq:32}
G_1'(\zeta_b^*,\zeta_a) = e^{-i\omega T s} \frac{(1 + \zeta_b^* \zeta_a e^{i\omega T})^{2s}}{(1+|\zeta_b|^2)^s(1+|\zeta_a|^2)^s} e^{\frac{i}{2}\omega T}
\end{equation}
which is incorrect
\begin{equation}\label{eq:33}
{\rm Tr}\{ G_1'\} = e^{\frac{i}{2}\omega T} \sum_{j=-s}^s e^{-i\omega T j}.
\end{equation}
To resolve the puzzle we ought to assume that the procedure described in Eq. (\ref{eq:21}), for the identification of the classical continuum Hamiltonian, does not pertain to the asymptotic limit only but it is valid for all spin values. What distinguishes each case is the canonical transformation that defines the classical ``momentum'' and
``position'' variables.

Thus, for the realization of the recipe (\ref{eq:21}) for finite $s$, and in accordance with Eqs. (\ref{eq:22a}) and (\ref{eq:22b}), we are led to define the classical variables
\begin{eqnarray}
\label{eq:34a} q &\equiv& \langle \zeta|\frac{\hat{\alpha}^\dag + \hat{\alpha}}{\sqrt{2}}|\zeta\rangle,\\
\label{eq:34b} p &\equiv& i \langle \zeta|\frac{\hat{\alpha}^\dag - \hat{\alpha}}{\sqrt{2}}|\zeta\rangle.
\end{eqnarray}
Using the Holstein-Primakoff transformation and the fact that
\begin{eqnarray}
\label{eq:35a} \langle\zeta|\hat{S}_-|\zeta\rangle &=& \frac{2s}{1+|\zeta|^2}\zeta^*,\\
\label{eq:35b} \langle\zeta|\hat{S}_+|\zeta\rangle &=& \frac{2s}{1+|\zeta|^2}\zeta,
\end{eqnarray}
we immediately determine the classical variables (\ref{eq:34a}) and (\ref{eq:34b}) to have the form:
\begin{eqnarray}
\label{eq:36a} q &=& \frac{1}{\sqrt{2}} (\zeta^* + \zeta)\sqrt{\frac{2s}{1+|\zeta|^2}},\\
\label{eq:36b} p &=& \frac{i}{\sqrt{2}} (\zeta^* - \zeta)\sqrt{\frac{2s}{1+|\zeta|^2}}.
\end{eqnarray}
Note that at the asymptotic limit $s\rightarrow\infty$ the variables (\ref{eq:22a}) and (\ref{eq:22b}) are recovered.

Thus, according to our prescription, the Hamiltonian entering in the integral (\ref{eq:04}) is not the ``standard'' one as indicated in Eq. (\ref{eq:09}) but the one produced after the canonical change of variables (\ref{eq:36a}) and (\ref{eq:36b}) that defines the Feynman version of (\ref{eq:09}):
\begin{equation}\label{eq:37}
H_{1F} = \omega\left[ s-\frac{1}{2}(p^2 + q^2) + \frac{1}{2}\right]
= \omega \left( s- 2s\frac{|\zeta|^2}{1 + |\zeta|^2} + \frac{1}{2}\right).
\end{equation}
Repeating now the steps that led from Eq. (\ref{eq:10}) to Eq. (\ref{eq:11}) we find
\begin{eqnarray}
\nonumber G_1'(\zeta_b^*,\zeta_a) &=& e^{-i\omega T s} \frac{(1 + \zeta_b^* \zeta_a e^{i\omega T})^{2s} e^{\frac{i}{2}\omega T}}{(1+|\zeta_b|^2)^s(1+|\zeta_a|^2)^s} \times\\
\label{eq:38} && \times \mathop{\int \mathcal{D}^2 \eta} \limits_{\substack{\eta^*(T)=0 \\ \eta(0)=0}} e^{i\int_0^T d\tau \left(\frac{i}{2}(\dot{\eta}\eta^* - \dot{\eta}^*\eta) + \omega|\eta|^2\right)}.
\end{eqnarray}
Adopting the result (\ref{eq:26}) we get the correct answer.

The main conclusion of this section is that the use of the Feynman Hamiltonian instead of the standard one together with the symmetrized definition of the path integrals leads, without inconsistencies, to the correct result.

However, the example we used was very simple and we have to consider more
complicated Hamiltonians in order to check our proposal. In what follows we shall call ``standard'' the procedure that adopts the form $H(\zeta^*,\zeta) = \langle \zeta|\hat{H}|\zeta\rangle$ for the classical Hamiltonian and takes into consideration presence of the ``extra phase'' factor.

\section{\label{sec:III}A less trivial example}
In this section we shall consider the less trivial Hamiltonian $\hat{H}_2 = \omega\hat{S}_z^2$ and the correlator
\begin{equation}\label{eq:39}
G_2(\zeta_b^*,\zeta_a) = \langle\zeta_b| e^{-iT\hat{H}_2}|\zeta_a\rangle.
\end{equation}
Due to the simplicity of the system this amplitude can be exactly evaluated:
\begin{equation}\label{eq:40}
G_2(\zeta_b^*,\zeta_a)= \sum_{j=-s}^s e^{-i\omega T j^2} \langle\zeta_b|s,j\rangle\langle s,j|\zeta_a\rangle,
\end{equation}
with
\begin{equation}\label{eq:41}
\langle\zeta|s,j\rangle = \frac{1}{(1+|\zeta|^2)^s}\left[\frac{(2s)!}{(s-j)!(s+j)!}\right]^{1/2} \zeta^{s-j}.
\end{equation}
However, the attempt to get the result (\ref{eq:40}) using the standard rules for performing coherent- state path integration in the continuum, fails. As we shall confirm the calculation of ${\rm Tr} \left\{ e^{-iT\hat{H}_2}\right\}$ in the standard framework (with the inclusion of the extra phase) yields the result~\cite{Wilson11}
\begin{equation}\label{eq:42}
 {\rm Tr} \left\{ e^{-iT\hat{H}_2}\right\} =\sum_{j=-s}^s e^{-i\omega T j^2 + i\omega T\frac{j^2-s^2}{2s}},
\end{equation}
which is wrong for every finite $s$ except for $s=1/2$. The correct answer is recovered only at the asymptotic (classical) limit $s\rightarrow\infty$.

However, the calculation in the continuum can be successfully executed by following the proposal presented in the previous section the main ingredient of which is the identification of the classical Hamiltonian entering in the path integral representation of the amplitude (\ref{eq:39}). To this end we appeal, once again, to the Holstein-Primakoff transformation to write
\begin{equation}\label{eq:43}
\hat{H}_2 = \omega\left( s^2 - 2s\hat{\alpha}^\dag\hat{\alpha} + \hat{\alpha}^\dag\hat{\alpha}\hat{\alpha}^\dag\hat{\alpha}\right).
\end{equation}
Expressing the annihilation and creation operators in terms of the quadratures we find
\begin{equation}\label{eq:44}
\hat{H}_{2F} = \omega\left[ s^2 - 2s\left(\frac{\hat{p}^2 + \hat{q}^2}{2} - \frac{1}{2} \right) + \left(\frac{\hat{p}^2 + \hat{q}^2}{2} - \frac{1}{2} \right)^2 \right].
\end{equation}
According to our prescription to obtain the classical Hamiltonian we replace the ``position'' and ``momentum'' operators by their classical values (\ref{eq:36a}) and (\ref{eq:36b})
\begin{eqnarray}\label{eq:45}
 H_{2F} &=& \omega \left[ s^2 - 2s\left(\frac{p^2 + q^2}{2} - \frac{1}{2} \right) + \left(\frac{p^2 + q^2}{2} - \frac{1}{2} \right)^2 \right] \\
\nonumber &=& \omega\left( s^2 +s + \frac{1}{4}\right) + \omega(2s)^2 \left( \frac{|\zeta|^2}{1+|\zeta|^2}\right)^2 - \omega 2s (2s+1)\frac{|\zeta|^2}{1+|\zeta|^2}.
\end{eqnarray}
It would be helpful at this point to write the form of the classical Hamiltonian had we adopted the ``standard'' prescription
\begin{equation}\label{eq:46}
H_2 = \langle\zeta|\hat{H}_2|\zeta\rangle = \omega s^2 + \omega 2s(2s-1) \left( \frac{|\zeta|^2}{1+|\zeta|^2}\right)^2 - \omega 2s(2s-1)\frac{|\zeta|^2}{1+|\zeta|^2}.
\end{equation}
Obviously the two expressions are the same only at the asymptotic limit $s\rightarrow\infty$. In the framework of our proposal the amplitude (\ref{eq:39}) assumes the form
\begin{eqnarray}
\nonumber G_{2F}(\zeta_b^*,\zeta_a) &=& \mathop{\int \mathcal{D} \mu(\zeta)} \limits_{\substack{\zeta^*(T)=\zeta_b^* \\ \zeta(0)=\zeta_a}} e^{\gamma_{ba}(\zeta^*,\zeta)
+ i\int_0^T d\tau \left(is\frac{\dot{\zeta}\zeta^* - \dot{\zeta}^*\zeta}{1+|\zeta|^2} - H_{2F}(\zeta^*,\zeta) \right)} \\
\nonumber &=& e^{-i\omega T \left(s^2 + s + 1/4\right)} \mathop{\int \mathcal{D} \mu(\zeta)} \limits_{\substack{\zeta^*(T)=\zeta_b^* \\ \zeta(0)=\zeta_a}} \exp\Bigg\{\gamma_{ba}(\zeta^*,\zeta) +\\
\nonumber && i\int_0^T d\tau \left(is\frac{\dot{\zeta}\zeta^* - \dot{\zeta}^*\zeta}{1+|\zeta|^2} \right. \\
 && \left. - \omega(2s)^2 \left(\frac{|\zeta|^2}{1+|\zeta|^2}\right)^2 + \omega 2s(2s+1)\frac{|\zeta|^2}{1+|\zeta|^2}\right)\Bigg\}.\label{eq:47}
\end{eqnarray}
To proceed we shall use the Hubbard-Stratonovich~\cite{Stratonovich57,Hubbard59,Muhlshegel,Halpern,Jevicki81} transformation which in our case can be realized through the introduction of the collective field $\rho(\tau) = |\zeta|^2/(1+|\zeta|^2)$. This can be consistently achieved~\cite{Kordas14} by making use of the functional identity
\begin{eqnarray}
\nonumber 1 &=& \int\mathcal{D}\rho~\delta\left[ \rho - \frac{|\zeta|^2}{1+|\zeta|^2} \right]\\
\label{eq:48}&=& \int\mathcal{D}\rho\int\mathcal{D}\sigma e^{-i\int_0^T d\tau \left(\rho - \frac{|\zeta|^2}{1+|\zeta|^2}\right)\sigma}.
\end{eqnarray}
Combining Eqs. (\ref{eq:47}) and (\ref{eq:48}) we can rewrite the correlator in the following form:
\begin{eqnarray}
\nonumber G_{2F}(\zeta_b^*,\zeta_a) &=& e^{-i\omega T\left(s^2 + s + \frac{1}{4}\right)} \int\mathcal{D}\rho\int\mathcal{D}\sigma \exp\Bigg\{ -i\omega\int_0^T d\tau ((2s)^2 \rho^2 \\
\label{eq:49} && - 2s(s+1)\rho) - i\int_0^T d\tau \rho\sigma\Bigg\} \tilde{G}(\zeta_b^*,\zeta_a;\sigma),
\end{eqnarray}
where in this expression we wrote
\begin{equation}\label{eq:50}
\tilde{G}(\zeta_b^*,\zeta_a;\sigma) = \mathop{\int \mathcal{D} \mu(\zeta)} \limits_{\substack{\zeta^*(T)=\zeta_b^* \\ \zeta(0)=\zeta_a}}
e^{\gamma_{ba}(\zeta^*,\zeta) + i\int_0^T d\tau \left(is\frac{\dot{\zeta}\zeta^* - \dot{\zeta}^*\zeta}{1+|\zeta|^2} + \sigma\frac{|\zeta|^2}{1+|\zeta|^2} \right)}.
\end{equation}
This integral has the same structure as the integral in Eq. (\ref{eq:10}) (with the change $\omega\rightarrow\sigma/2s$). Thus the calculation is straightforward and provided that we adopt the symmetrized discrete definition of the functional integral we get:
\begin{eqnarray}\label{eq:51}
 \tilde{G}_S(\zeta_b^*,\zeta_a;\sigma) &=& \frac{\left(1+\zeta_b^*\zeta_a e^{\frac{i}{2s}\int_0^T d\tau\sigma}\right)^{2s} e^{\frac{i}{4s}\int_0^T d\tau\sigma}}{(1+|\zeta_b|^2)^s (1+|\zeta_a|^2)^s} \\
\nonumber &=& \frac{\sum_{p=0}^{2s} \left(\begin{array}{c}
2s \\ p
\end{array}\right) (\zeta_b^*\zeta_a)^p e^{\frac{i}{2s}\left(p+\frac{1}{2}\right) \int_0^T d\tau\sigma} }{(1+|\zeta_b|^2)^s (1+|\zeta_a|^2)^s}.
\end{eqnarray}
Inserting this result into Eq. (\ref{eq:49}) we find:
\begin{eqnarray}
\nonumber G_{2F}(\zeta_b^*,\zeta_a) &=& \frac{e^{-i\omega T \left(s^2 + s + \frac{1}{4}\right)}}{(1+|\zeta_b|^2)^s (1+|\zeta_a|^2)^s}\times \\
\nonumber && \times\sum_{p=0}^{2s} \left(\begin{array}{c}
2s \\ p
\end{array}\right) (\zeta_b^* \zeta_a)^p \times \\
\nonumber && \times \int\mathcal{D}\rho e^{-i\omega \int_0^T d\tau ((2s)^2\rho^2 - 2s(s+1)\rho)}\times \\
\label{eq:52}&& \times\int\mathcal{D}\sigma e^{-i\int_0^T d\tau \left(\rho - \frac{1}{2s}\left(p+\frac{1}{2}\right)\right)\sigma}.
\end{eqnarray}
The last integral results to a delta function instructing that $\rho = (1/2s)(p+1/2)$. Thus the integral over $\rho$ is immediately performed yielding the exact result:
\begin{eqnarray}
\nonumber G_{2F}(\zeta_b^*,\zeta_a) &=& \frac{e^{-i\omega T \left(s^2 + s + \frac{1}{4}\right)}}{(1+|\zeta_b|^2)^s (1+|\zeta_a|^2)^s}\times \\
\nonumber && \times\sum_{p=0}^{2s} \left(\begin{array}{c} 2s \\ p \end{array}\right) (\zeta_b^* \zeta_a)^p e^{-i\omega T(p^2-2sp)}\\
\nonumber &=& \sum_{j=-s}^s e^{-i\omega T j^2} \frac{(2s)!}{(s-j)!(s+j)!}\times\\
\nonumber && \times\frac{(\zeta_b^* \zeta_a)^{s-j}}{(1+|\zeta_b|^2)^s (1+|\zeta_a|^2)^s}\\
\label{eq:53}&=& \sum_{j=-s}^s e^{-i\omega T j^2} \langle \zeta_b|s,j\rangle\langle s,j|\zeta_a\rangle.
\end{eqnarray}
For comparison we can repeat the previous steps beginning with the standard Hamiltonian (\ref{eq:46}). In such a case the expression (\ref{eq:49}) assumes the form:
\begin{eqnarray}
\nonumber G_2(\zeta_b^*,\zeta_a) &=& e^{-i\omega Ts^2} \int\mathcal{D}\rho \int\mathcal{D}\sigma e^{-i\omega 2s(s+1)\int_0^T d\tau (\rho^2-\rho)}\\
\label{eq:54} && \times e^{-i\int_0^T d\tau\rho\sigma} \tilde{G}(\zeta_b^*,\zeta_a;\sigma).
\end{eqnarray}
Following the standard procedure for the evaluation of the last integral and taking into account the ``extra phase'' contribution we find that:
\begin{eqnarray}\label{eq:55}
 \tilde{G}(\zeta_b^*,\zeta_a;\sigma) &=& \frac{\left(1+\zeta_b^*\zeta_a e^{\frac{i}{2s}\int_0^T d\tau\sigma}\right)^{2s}}{(1+|\zeta_b|^2)^s (1+|\zeta_a|^2)^s} \\
\nonumber &=& \frac{\sum_{p=0}^{2s} \left(\begin{array}{c}
2s \\ p
\end{array}\right) (\zeta_b^*\zeta_a)^p e^{\frac{i}{2s} p \int_0^T d\tau\sigma} }{(1+|\zeta_b|^2)^s (1+|\zeta_a|^2)^s}.
\end{eqnarray}
Once again the integral over $\sigma$ forces $\rho=p/2s$ and the amplitude (\ref{eq:54}) reads
\begin{eqnarray}
\nonumber G_2(\zeta_b^*,\zeta_a) &=& \frac{e^{-i\omega Ts^2}}{(1+|\zeta_b|^2)^s (1+|\zeta_a|^2)^s} \sum_{p=0}^{2s} \left(\begin{array}{c}
2s \\ p \end{array}\right) (\zeta_b^*\zeta_a)^p e^{-i\omega T\frac{2s-1}{2s}(p^2-2sp)}\\
\label{eq:56} &=& \sum_{j=-s}^s e^{-i\omega T j^2 + i\omega T\frac{j^2-s^2}{2s}} \langle \zeta_b|s,j\rangle \langle s,j|\zeta_a\rangle.
\end{eqnarray}
As the comparison with the exact formula (\ref{eq:40}) proves, this is a wrong result. If we take, for example, the case $s=1$ we get for the trace of the amplitude (\ref{eq:56}) the result already indicated in~\cite{Wilson11}
\begin{equation}\label{eq:57}
{\rm Tr} G_2 = 2e^{-i\omega T} + e^{-i\omega T/2} \neq {\rm Tr} e^{-iT\hat{H}_2}.
\end{equation}

At this point it would be enlightening to summarize our proposal for constructing time continuous spin coherent state path integrals:

Suppose that the dynamics of a spin system is described by a quantum Hamiltonian of the form $\hat{H}=\hat{H}(\hat{S}_x,\hat{S}_y,\hat{S}_z)$. The first step is to make use of the Holstein-Primakoff transformation (\ref{eq:20}) to rewrite the Hamiltonian in terms of the bosonic creation and annihilation operators: $\hat{H}=\hat{H}(\hat{\alpha}^\dag,\hat{\alpha})$. Next come the introduction of the quadratures $\hat{q},\hat{p}$ through the relations $\hat{\alpha}=(\hat{q}+i\hat{p})\sqrt{2},~\hat{\alpha}^\dag=(\hat{q}-i\hat{p})\sqrt{2}$. This step yields the
recasting of the Hamiltonian: $\hat{H}=\hat{H}(\hat{p},\hat{q})$. The crucial step is the third one consisting of the replacement of the quantum Hamiltonian by its classical counterpart $\hat{H}\rightarrow H_F(p,q)$, where $p$ and $q$ are the representation of the quadratures in the coherent state basis: $p=\langle\zeta|\hat{p}|\zeta\rangle,~q=\langle\zeta|\hat{q}|\zeta\rangle$. As the analysis dictates, the resulting classical Hamiltonian must be used to define the continuum action that weighs the summation of paths in the manifold we are working with.

Note that the above recipe does not depend on the value of the spin of the system. This is the reason the above described road of Hamiltonian construction, is essentially the same for spin and bosonic systems as the latter are the asymptotic limit of the former. Roughly speaking, the key idea is to begin from the well-defined phase-space Feynman path integral, and by making suitable canonical transformations to arrive at the description of the problem we are interested in. This procedure also fixes the permissible discrete definition of the path integration.

In the exactly solvable example discussed above, the final calculation has been reduced to that of a simple harmonic oscillator. However, this is not neither the general case nor the most interested one. On the contrary, the coherent-state path integration has been commonly used for semiclassical calculations. While the structure of this kind of calculations remain the same in our approach, higher order differences arise due to the different classical action entering into the path integrals. As a concrete example consider the Lipkin-Meshkov-Glik (LMG) model~\cite{Lipkin65} in which the tunneling splitting has been quasi-classically calculated~\cite{Garg03}:
\begin{equation}
\hat{H} = \frac{w}{\sqrt{2}(2s-1)}(\hat{S}_+^2 + \hat{S}_-^2) + \frac{sw}{\sqrt{2}},~~w>0.
\end{equation}
The ``standard" form of this Hamiltonian reads as follows:
\begin{equation}
\langle\zeta|\hat{H}|\zeta\rangle = \sqrt{2}sw\frac{\zeta^{*2} +\zeta^2}{(1+|\zeta|^2)^2} + \frac{sw}{\sqrt{2}}.
\end{equation}
However, our recipe leads to the following form of the relevant classical Hamiltonian:
\begin{equation}
H_F(\zeta^*,\zeta) = \sqrt{2}\frac{2s^2}{2s-1} w \frac{\zeta^{*2} +\zeta^2}{(1+|\zeta|^2)^2}\left(1 + \frac{1+|\zeta|^2}{2s}\right) + \frac{sw}{\sqrt{2}}.
\end{equation}
The two expressions coincide only at the asymptotic limit $s\rightarrow\infty$ thus we expect our approach to be more accurate when corrections of order $1/s$ are significant. In a forthcoming work we shall present detail results on the subject.

\section{\label{sec:IV}Conclusions}
In this work we present a method for defining and handling time-continuous spin coherent-sate path integral without facing inconsistencies. Such a path-integral formulation opens new possibilities for applying semiclassical approximations with improved accuracy in quantum spin models. Moreover, this formalism can be also applied in bosonic systems, such as the two-site Bose-Hubbard model, $\hat{H}_{BH} = -J\hat{S}_x + U\hat{S}_z^2$ at the large $N$ limit~\cite{Legget01,Holthaus01,Diaz12}, which are of increasing interest both theoretically and experimentally. The aim of this paper is not the presentation of new results. It is, rather, an attempt to set a solid basis of reliable calculations.

Our approach is based on two pylons. The first is the adoption of a discretized form of the path integrals that is invariant under canonical transformations. The second is the three simple steps for identifying the Hamiltonian that weighs the paths in the non-flat manifold $\bar{C}^1$. In the first step we use the Holstein-Primakoff transformation to rewrite the quantum Hamiltonian in terms of ``position'' and ``momentum'' operators. The second step consists of constructing the Feynman phase-space integral in which the classical version of the Hamiltonian enters. The third step is a canonical change of variables that produces the form
of the Hamiltonian which enters into the time-continuous spin coherent-state path integral. We have followed this simple method to derive, directly in the continuum, the correct result for the simple case $\hat{H}\sim \hat{S}_z^2$ for which inconsistencies have been repeatedly reported.

{}

\end{document}